\documentclass{SciPost}

\hypersetup{
    colorlinks,
    linkcolor={red!50!black},
    citecolor={blue!50!black},
    urlcolor={blue!80!black}
}

\usepackage[bitstream-charter]{mathdesign}
\DeclareSymbolFont{usualmathcal}{OMS}{cmsy}{m}{n}
\DeclareSymbolFontAlphabet{\mathcal}{usualmathcal}

\fancypagestyle{SPstyle}{
\fancyhf{}
\lhead{\colorbox{scipostdeepblue}{\bf \color{white} ~SciPost Physics Proceedings }}
\rhead{{\bf \color{scipostdeepblue} ~Submission }}

\fancyfoot[C]{\textbf{\thepage}}
}

\begin{document}

\pagestyle{SPstyle}

\begin{center}{\Large \textbf{\color{scipostdeepblue}{
Modern Machine Learning and Particle Physics \\ Phenomenology at the LHC
}}}\end{center}

\begin{center}\textbf{
Maria Ubiali \textsuperscript{1*}
}\end{center}

\begin{center}
{\bf 1} DAMTP, University of Cambridge, Wilberforce Road, Cambridge CB3 0WA, UK
\\[\baselineskip]
$\star$ \href{mailto:M.Ubiali@damtp.cam.ac.ukj}{\small M.Ubiali@damtp.cam.ac.uk}
\end{center}

\definecolor{palegray}{gray}{0.95}
\begin{center}
\colorbox{palegray}{
  \begin{tabular}{rr}
  \begin{minipage}{0.37\textwidth}
    \includegraphics[width=60mm]{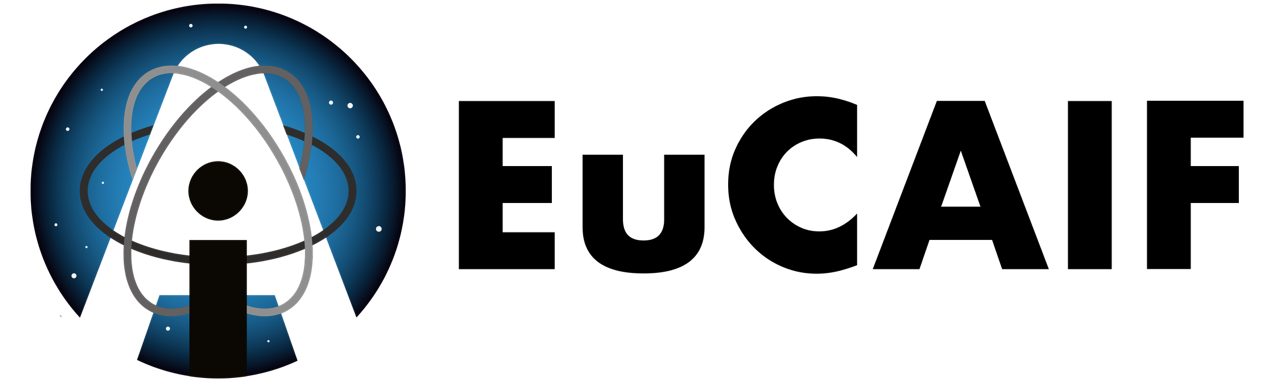}
  \end{minipage}
  &
  \begin{minipage}{0.5\textwidth}
    \vspace{5pt}
    \vspace{0.5\baselineskip} 
    \begin{center} \hspace{5pt}
    {\it The 2nd European AI for Fundamental \\Physics Conference (EuCAIFCon2025)} \\
    {\it Cagliari, Sardinia, 16-20 June 2025
    }
    \vspace{0.5\baselineskip} 
    \vspace{5pt}
    \end{center}
    
  \end{minipage}
\end{tabular}
}
\end{center}

\section*{\color{scipostdeepblue}{Abstract}}
\textbf{\boldmath{%
Modern machine learning is driving a paradigm shift in particle physics phenomenology at the 
Large Hadron Collider. This short review examines the transformative role of machine learning 
across the entire theoretical prediction pipeline, from parton-level calculations to full  
simulations. We discuss applications to scattering amplitude computations, phase space integration, 
Parton Distribution Function determination, and parameter extraction. Some critical frontiers  
for the field including uncertainty quantification, the role of symmetries, and interpretability are 
highlighted.
}}

\vspace{\baselineskip}

\noindent\textcolor{white!90!black}{%
\fbox{\parbox{0.975\linewidth}{%
\textcolor{white!40!black}{\begin{tabular}{lr}%
  \begin{minipage}{0.6\textwidth}%
    {\small Copyright attribution to authors. \newline
    This work is a submission to SciPost Phys. Proc. \newline
    License information to appear upon publication. \newline
    Publication information to appear upon publication.}
  \end{minipage} & \begin{minipage}{0.4\textwidth}
    {\small Received Date \newline Accepted Date \newline Published Date}%
  \end{minipage}
\end{tabular}}
}}
}


\section{Introduction}
\label{sec:intro}
The discovery of the Higgs boson at the LHC in 2012 inaugurated a new era in high-energy physics, 
opening the exploration of the new Yukawa force and confronting persistent puzzles that the Standard Model (SM) 
cannot explain, including dark matter, neutrino masses, matter-antimatter asymmetry, and the nature of gravity. 
The LHC is rapidly evolving into a precision machine, able to measure small deviations from SM predictions. 
This precision programme demands rigorous methodologies to interpret the extremely accurate and highly correlated 
experimental data and the increasingly complex theoretical predictions.

Particle physics phenomenology serves as the bridge between experimental observations and theoretical models, 
connecting what is measured at colliders to our fundamental understanding of Nature. In Bayesian terms, 
we seek the probability of a given theory given experimental data. At the LHC, however, the likelihood 
itself is extraordinarily complex, requiring a {\it divide-and-conquer} approach across multiple stages 
of the theoretical computations, from a given Lagrangian -- whether the SM one or the one of some beyond-SM (BSM) 
scenarios. 

From this starting point, predictions flow through several distinct levels: parton-level calculations involving 
scattering amplitudes, phase space integration, and Parton Distribution Functions (PDFs); particle-level simulations 
including QCD and QED radiation through parton showers, fragmentation, and hadronisation; and finally detector-level 
modeling encompassing detector simulation, event reconstruction, and selection criteria.

Machine Learning (ML) is revolutionising every component of this intricate chain. 
The HEP ML living review~\cite{hepmllivingreview} documents the breadth of 
this transformation. This short review starts with an attempt in Sect.~\ref{sec:2} 
to provide an overview of some selected applications where ML provides unique advantages, 
focussing on the ingredients entering parton level predictions. 
In Sect.~\ref{sec:3} we provide a brief sketch on some 
applications that enable to go beyond parton level, and simulate the full event. Finally, 
before concluding, in Sect.~\ref{sec:4} we highlight some of the most compelling challenges 
in the field. 

\section{Machine Learning for Parton-Level Predictions}
\label{sec:2}
A differential cross section at the parton level at the LHC can be expressed as the product of squared 
scattering amplitudes, $n$-particle phase space, and PDFs characterising the 
proton's subnuclear structure. In this section we provide some examples that show how ML enables to 
extend, improve, simplify and speed up the determination of the main ingredients that 
enter any parton-level theoretical predictions at the LHC, namely the calculation 
of scattering amplitudes (Sect.~\ref{sec:21}), the phase space integration (Sect.~\ref{sec:22}) 
and the PDFs (Sect.~\ref{sec:23}). 
Finally in Sect.~\ref{sec:24} we will quickly discuss paramater determination from parton 
level predictions. The list of applications mentioned here is certainly far from complete, 
and it is subject to the auhor's knowledge and unavoidable bias. Nevertheless we hope it gives 
an idea of the great power of machine learning in tackling the problems that we face as particle 
physicists. 

\subsection{Scattering Amplitude Calculations}
\label{sec:21}
Computing scattering amplitudes, particularly at higher multiplicities and loop orders, represents one 
of the most computationally intensive tasks in theoretical physics.
Machine learning offers a natural solution through regression problems that exploit neural network (NN) 
flexibility to accelerate the computation of amplitudes. The approach trains ML regressors, often NNs or ensembles of NNs, 
on pre-computed "true" amplitudes evaluated at numerous phase space points. These trained networks then predict 
amplitudes accurately and rapidly for new phase space configurations. 

Recent applications span processes from simple two-to-two scattering to complex multi-particle final states, 
with NNs demonstrating performance superior to traditional numerical simulations for higher-multiplicity 
processes~\cite{Bishara:2019iwh,Badger:2020uow,Buckley:2020bxg,Bury:2020ewi,Sombillo:2021rxv,
Aylett-Bullock:2021hmo,Danziger:2021eeg,Alnuqaydan:2022ncd,Maitre:2023dqz,Janssen:2023ahv,Herrmann:2025nnz}.
For ML surrogates to generate reliable higher-order predictions, they must achieve precision 
reflecting underlying theoretical accuracy. Comprehensive uncertainty estimation becomes crucial 
for using these surrogates in actual simulations. Several complementary approaches address 
this challenge.
Heteroscedastic losses incorporate unknown uncertainties directly into the loss function, 
learning them through deterministic networks. Bayesian neural networks on the other hand provide natural uncertainty 
quantification through posterior distributions over network parameters. Repulsive ensembles offer an 
alternative framework for tracking both statistical and systematic uncertainties. Benchmarking studies 
compare these methods' performance in representative amplitude calculations, see 
for example Refs.~\cite{Badger:2022hwf,Bahl:2024gyt,Bahl:2025xvx}.

Beyond amplitude evaluation itself, ML techniques can be used to enhance the computation of multi-loop integrals 
that appear at next-to-leading-order (NLO) and higher order corrections. These integrals typically contain integrable 
singularities on the real axis, necessitating contour deformation into the complex plane. NN-assisted 
algorithms based on normalising flows significantly amplify the precision of standard contour deformation 
methods~\cite{Winterhalder:2021ngy}. 
Alternative approaches employ NNs to solve numerically the differential equations satisfied by Feynman 
integrals, offering complementary strategies for multi-loop calculations~\cite{Calisto:2023vmm,Maitre:2022xle}.

A somewhat different and exciting direction is to use ML to simplify algebraic expressions. 
Integration of scattering amplitudes generates mathematical functions lacking classical 
simplification algorithms, such as for example generalised polylogarithms, 
In ~\cite{Dersy:2022bym} reinforcement learning is used by treating known identities as 
moves in a game, training agents to apply these transformations. Alternatively transformer networks 
are used to translate complicated expressions into simplified forms, essentially performing symbolic 
manipulation through learned patterns. 
Similar techniques using transformers and contrastive learning simplify spinor-helicity 
representations of scattering amplitudes~\cite{Cheung:2024svk}.
In the context of planar N=4 Super Yang-Mills theory, a further ML application appears quite exciting, 
as transformer models are used to predict coefficients at loop 
order $L$ using only small subsets of related coefficients from loop order$L-1$, demonstrating genuine pattern 
recognition across loop levels~\cite{Cai:2024znx,Cai:2025atc}.

\subsection{Phase Space Integration}
\label{sec:22}
Another critical computational bottleneck in the computation of parton-level cross sections is the 
efficient integration of squared amplitudes over the phase space.  
The challenge lies in importance sampling and multi-channel techniques whose efficiency depends on 
variable transformations and channel selection. The traditional VEGAS~\cite{Lepage:1977sw} algorithm implements adaptive importance 
sampling by fitting bins with equal probability and varying width. While computationally cheap, VEGAS neglects 
correlations between variables and struggles with multimodal functions when peaks misalign with coordinate axes. 

Bijective normalising flows provide a powerful ML-based alternative. These architectures put together invertible, 
learnable transformations with exact likelihood evaluation through change-of-variables formulas. 
By redistributing input random variables through learned mappings, normalising flows adapt naturally 
to complex integrand structures. Applications to phase space integration demonstrate substantial 
improvements over traditional approaches, see for example~\cite{Muller:2018pvg,Bothmann:2020ywa,
Gao:2020vdv,Gao:2020zvv,Chen:2020nfb,Pina-Otey:2020hzm,Deutschmann:2024lml}.
The MadNIS framework~\cite{Heimel:2022wyj,Heimel:2023ngj,Heimel:2024wph}  
exemplifies the practical integration of ML tools with established tools, as it 
combines the standard automated event generator MadGraph~\cite{Alwall:2014hca} with two NN components: a 
channel-weight network encoding local multi-channel weights, and an invertible network implementing 
normalising flows that function efficiently in both forward and inverse directions. This hybrid approach 
achieves in some cases a strong improvement in both accuracy and efficiency compared to standard methods, 
while maintaining compatibility with existing computational workflows.

\subsection{Parton Distribution Functions}
\label{sec:23}
PDFs encode the probability that a parton carries a given 
momentum fraction of its parent proton. These universal, non-perturbative objects cannot 
be computed from first principles and must be extracted from experimental data~\cite{Amoroso:2022eow,Ubiali:2024pyg}.  
The PDF fitting problem constitutes an infinite-dimensional inverse problem~\cite{DelDebbio:2021whr}: given a finite 
set of discrete experimental measurements, determine $n_{\rm flac}$ continuous functions of $x$ 
with proper uncertainty estimate. The NNPDF collaboration pioneered a NN approach to this challenge~\cite{Ball:2008by}, 
using NNs as flexible functional parametrisation that did not introduce parametrisation bias that 
might affect traditional polynomial parametrisations. 
The NNPDF4.0 methodology is the modern version of the initial NNPDF idea 
and it employs a single deep neural network with hyperparameters optimized through $K$-fold cross-validation 
procedures~\cite{Carrazza:2019mzf,NNPDF:2021njg}. 
Approximately five thousand parton-level data points from electron-proton, electron-nucleon, 
proton-antiproton, and proton-proton experiments are used to extract eight PDFs. 
Uncertainty estimation follows Monte Carlo error propagation principles through bootstrap 
sampling~\cite{Costantini:2024wby}. The procedure generates pseudo-data replicas by adding 
random fluctuations to experimental measurements according to their multivariate normal distributions 
with experimental covariance matrices. For each replica, training with validation-based stopping 
yields optimal parameter values. Repeating this process provides importance sampling of the posterior 
distribution in PDF space, projecting from data-space sampling. 
This deep learning-based methodology yields smaller uncertainties in data regions 
while maintaining conservative extrapolation uncertainties~\cite{Chiefa:2025loi}.
The NNPDF4.0 public code~\cite{NNPDF:2021uiq} enables testing and reproducibility, 
with methodology rigorously scrutinised through statistical closure tests~\cite{Barontini:2025lnl}.
Another ML application in the context of NNPDF is the use of generative adversarial networks 
providing compressed PDF representations with fewer replicas matching full 
ensemble accuracy~\cite{Carrazza:2015hva,Carrazza:2021hny}.
The techniques can now be tested in a fully automated and flexible PDF regression 
framework {\tt Colibri}~\cite{Costantini:2025agd}, in which the NN parametrisation 
and the bootstrap uncertainty estimate can be tested against other models and Bayesian error propagation,  
assessing the reliability of ML approaches for this foundational task.

\subsection{Parameter determination}
\label{sec:24}
Another key challenge at the LHC is the precise and robust determination of SM and BSM parameters 
from LHC data. Determining which 
measurements provide optimal sensitivity to specific parameters, and selecting  
observables and binnings, challenges human two-dimensional intuition in high-dimensional 
parameter spaces.
ML tools naturally "see" in arbitrarily high dimensions, identifying optimal variables 
and binning. Neural simulation-based inference exemplifies this capability~\cite{Brehmer:2020cvb,Bahl:2024meb}. 

Among the many applications that could be mentioned, an interesting one is the 
application of such techniques to the determination of 
the Wilson coefficients of the SM effective field theory (SMEFT). The SMEFT 
provides a model-independent framework for parametrising deviations 
from SM predictions through Wilson coefficients multiplying higher-dimensional 
operators~\cite{Brivio:2017vri}. 
As experimental precision and theoretical accuracy improve, identifying patterns 
of small deviations -- corresponding to non-zero Wilson coefficients -- would 
allow to extract hints on the underlying new physics model. 
In~\cite{GomezAmbrosio:2022mpm} unbinned parton-level likelihoods are generated and 
for each point of the phase space the SMEFT is compated to the SM, by training NN to approximate 
likelihood ratios. 
These trained classifiers compute signal strengths and enable inference about Wilson coefficients 
directly from unbinned, multidimensional data.
Related methodologies include parametrised classifiers for 
SMEFT~\cite{Brehmer:2019xox,Brehmer:2018eca,Brehmer:2018kdj,Chen:2020mev},  
tree-boosting approaches for learning EFT likelihoods~\cite{Chatterjee:2022oco}, 
and techniques for designing optimal observables through deep learning~\cite{Chen:2023ind,Long:2023mrj}. 
Recent ATLAS analyses employ these methods for measuring off-shell Higgs boson couplings to $Z$ 
bosons~\cite{ATLAS:2024jry,ATLAS:2025clx}, 
demonstrating practical deployment in experimental measurements.

Further developments related to the determination of SMEFT coefficients 
have to do with simultaneous fits of these parameters and 
PDFs using deep NNs~\cite{Costantini:2024xae,Iranipour:2022iak}, recognising 
the interplay and accounting for correlations between these quantities~\cite{Hammou:2023heg}. 

\section{Beyond Parton Level: End-to-End Simulations}
\label{sec:3}
In this section we briefly mention some of the numerous applications of ML to 
go beyond parton level, to a full simulation. Traditional Monte Carlo event generation 
proceeds sequentially through parton shower, hadronisation, detector simulation, and 
particle reconstruction stages. Each step introduces approximations and computational costs. 
End-to-end machine learning surrogates for fast event simulation learn multiple stages 
simultaneously, potentially improving both speed and robustness, see~\cite{Butter:2022rso} for 
a comprehensive review. 

Early approaches employed generative adversarial networks and variational autoencoders 
for calorimeter shower simulation and jet 
generation~\cite{Otten:2019hhl,Hashemi:2019fkn,DiSipio:2019imz,Butter:2019cae,ArjonaMartinez:2019ahl,Alanazi:2020klf}. 
While demonstrating proof-of-concept, these methods faced challenges with training stability and mode coverage. 
Normalising flows improved both speed and efficiency through their tractable likelihood evaluation 
and stable training dynamics~\cite{Stienen:2020gns,Butter:2021csz,Kach:2022qnf,Bellagente:2021yyh}. 
Recent diffusion models and transformer architectures achieve a very high level of precision in generating 
realistic detector-level events~\cite{Leigh:2023toe,Butter:2023fov,Leigh:2023zle}.
Importantly, conditional generative adversarial networks and transformer architectures 
enable inversion of the simulation chain, mapping from detector-level observations 
back to parton-level configurations. This, combined with uncertainty quantification 
through Bayesian NNs and classifier-based methods, provides end-to-end error control 
across the simulation pipeline~\cite{Nachman:2023clf}.
An alternative approach, optimal transport-based unfolding and simulation (OTUS), employs probabilistic 
autoencoders to learn bidirectional mappings between parton and reconstruction levels without 
requiring paired event samples. This framework has the potential to handle both simulation and unfolding, 
learning both directions simultaneously~\cite{Howard:2021pos}.

Something that is worth mentioning in this context is the matrix element method (MEM), 
building full likelihoods using matrix elements from theory combined 
with transfer functions describing detector response. This approach allows direct 
inference of fundamental theory parameters from reconstructed events. 
Modern implementations employ normalising flows with transformers for transfer functions, 
classifier networks for acceptance probabilities, and generative networks for efficient 
Monte Carlo sampling, while incorporating direct theory input for parton-level event generation. 
This integration of traditional theoretical calculations with ML components exemplifies 
the synergy between established methods and modern tools~\cite{Heimel:2023mvw,Butter:2022vkj}.
However, while traditional analyses from reconstructed events to parton level necessarily lose information 
through binning and hand-crafted observable selection. Access to full likelihoods would enable 
unbinned, multivariate analyses with optimal information extraction. 
While the full likelihood remains intractable in realistic settings, combining ML components 
for detector modeling with theoretical inputs for parton-level calculations approximates this ideal, 
enabling more powerful inference than conventional approaches.

\section{Frontiers: Uncertainty, Symmetry, and Interpretability}
\label{sec:4}
In this section, we summarise what we believe are the most exciting and relevant frontiers 
at the interface of ML and theoretical particle physics, namely uncertainty quantification, 
the encoding of theoretical constrains and symmetries in the ML models and interpretability. 

As ML is used across all steps of the pipeline to build theoretical predictions, 
it is paramount to ensure that ML tools provide robust results with comprehensive uncertainty 
quantification. High-energy physics possesses unique potential to advance from deterministic 
ML to probabilistic frameworks, given the rigorous uncertainty treatment 
and statistical methodologies used in the field.
Multiple approaches to uncertainty quantification have emerged across different contexts: 
bootstrap sampling for PDF uncertainties, heteroscedastic losses for amplitude regression, 
Bayesian neural networks for posterior distributions, repulsive ensembles for diverse predictions, 
and posterior sampling for generative models. Understanding relationships between these methods, 
benchmarking their performance, analysing prior dependence, and developing appropriate statistical 
tests represents crucial ongoing work.
The transition from deterministic outputs to probabilistic predictions fundamentally changes how ML 
integrates with fundamental physics. Rather than single point predictions, 
probabilistic models provide full distributions over possible outcomes, naturally 
incorporating both aleatoric uncertainty from irreducible stochasticity and epistemic uncertainty 
from limited training data or model capacity.

The second frontier has to do with smart inductive bias. 
Physics laws fundamentally respect symmetries, such as Lorentz invariance, 
gauge symmetry, and other transformation properties defining quantum field theories. 
Incorporating known symmetries directly into ML architectures provides powerful inductive 
biases that improve generalisation, efficiency, and physical consistency.
As an example Lorentz-equivariant transformers encode Lorentz symmetry into 
network architecture to provide appropriate latent representations of phase 
space points~\cite{brehmer2023geometricalgebratransformer,matthes2023unifiedframeworkcontrastivelearning,
Spinner:2024hjm,Brehmer:2024yqw}. 
Applications span amplitude regression, event classification, and generation, demonstrating that 
symmetry-aware architectures consistently outperform symmetry-agnostic alternatives while 
guaranteeing physically consistent behavior.

Beyond incorporating known symmetries, ML offers exciting potential for discovering 
symmetries in data. Detecting previously unknown symmetries would signify fundamental principles 
manifesting as physical laws and selection rules. Classifiers and symmetry generative adversarial 
networks explore deep learning approaches to symmetry discovery, with applications ranging from 
identifying approximate symmetries in complex systems to uncovering hidden patterns in 
HEP data~\cite{Betzler:2020rfg,Krippendorf:2020gny,Barenboim:2021vzh}.

Finally, the third frontier is interpretability. As physicists we seek understanding, 
striving for simplicity and unity in natural laws. While numerical simulations and 
black-box predictions suffice for some applications, particularly when comprehensive 
uncertainty quantification provides statistical rigor, analytical understanding offers 
distinct advantages including superior extrapolation and conceptual clarity.
Symbolic regression bridges machine learning and analytical formulas, learning 
complex functions from high-dimensional data while expressing results in closed form. 
There is an increasing number of Applications to 
HEP~\cite{Butter:2021rvz,Lu:2022joy,Tsoi:2024pbn,Morales-Alvarado:2024jrk,Dotson:2025omi,Makke:2025zoy,Bendavid:2025urn}, 
most using evolutionary algorithms that evaluate and optimize symbolic 
expressions such as {\tt PySR}~\cite{cranmer2023interpretablemachinelearningscience}. 
The resulting formulas provide both computational efficiency and physical insight unavailable 
from purely numerical approaches.
Complementary directions investigate understanding deep neural networks themselves using 
principles from quantum field theory or cosmological dynamics, see for 
example~\cite{Gukov:2024buj,Halverson:2024hax,Krippendorf:2022hzj}. These theoretical frameworks 
for analysing neural network behavior may reveal why certain architectures succeed and guide future development.

\section{Conclusion}
A revolution in LHC physics through modern machine learning is ongoing. 
Contemporary ML tools, combining regression, classification, generation, 
and conditional generation, benefit every stage of theoretical predictions at the LHC. 
The exceptional quality and quantity of high-energy physics data, with careful control of systematic 
uncertainties and correlations, provides labeled and well-characterised datasets that make particle 
physics an ideal testing ground for ML methodologies.
Critical frontiers for future development include comprehensive uncertainty quantification 
across all applications. 
The synergy between traditional theoretical methods and modern ML tools 
promises continued rapid progress. By maintaining physics rigour while embracing 
computational advances, the field strives toward the ultimate goal: extracting 
maximal information from collider data to reveal nature's fundamental principles.
As Alan Turing observed, "We can only see a short distance ahead, but we can see 
plenty there that needs to be done." The landscape of ML applications 
in particle physics phenomenology presents abundant opportunities for meaningful 
contributions that will shape the future of high-energy physics at the LHC and beyond.

\section*{Acknowledgements}
I am grateful to Stefano Forte, Ramon Winterhalten, Sven Krippendorf, Raquel Ambrosio-Gomez 
and Veronica Sanz for their precious input in preparing the talk. I thank the organisers for having 
invited me, as they helped me expand my horizon beyond the topics I was familiar with.

\paragraph{Funding information}
M.~U, is supported by the European Research Council under the European Unions Horizon 2020
research and innovation Programme (grant agreement n.950246), and partially by the STFC consolidated
grant ST/X000664/1.









\end{document}